# Sustainability of Scale-Free Properties in Synchronizations of Dynamic Scale-Free Networks

Rakib Hassan Pran*
ORCID iD: 0000-0002-6736-3741
*M.Sc. in Applied Statistics with Network Analysis,
International Laboratory for Applied Network Research,
National Research University Higher School of Economics,
Moscow, Russia

*Abstract*— Scale-free networks are ubiquitous in social, biological and technological networked systems. Dynamic Scale-free networks and their synchronizations are important to understand and predict the behavior of social, biological and technological networked systems. In this research, computational experiments have been conducted to understand the sustainability of scale-free properties during the time of synchronizations in dynamic scale-free networks. Two synchronization phenomena which are synchronization based on states of nodes with coupling configuration matrix and synchronization based on states of nodes with network centralities have been implemented for the synchronization in dynamic scale-free networks. In experiments, dynamic scale-free networks have been generated with a network generation algorithm and analyzed to understand the fluctuation from the scale-free properties in their phases during the time of synchronizations.

## I. Introduction

Scale-free networks are fundamental to study social, biological and technological networked systems [1]. To study scale-free networks, it is vital to understand the attributes of scale-free properties [1]. The term "scale-free" came from the branch of statistical physics called "the theory of phase transitions" which extensively explored power law distribution [1]. The scale-free network is scale-free because when we randomly choose a node from a scale-free network, we don't know what to expect such as the selected node's degree can be very small or arbitrarily large [1]. Scale-free networks do not have meaningful internal scale but have "scale-free" properties [1]. In existing literature, most of the scale-free properties are following power-law distribution for degree distribution in scale-free networks [1]. Along with power-law degree distribution, correlation analysis among applied social network measures into scale-free networks also existed in literature [10-11].

Besides the scale-free properties, it is also necessary to follow up the synchronization of evolving scale-free networks over time or phases for better understanding and predicting the behavior of social, biological and technological networked systems [2-4]. Various synchronization phenomena have been observed in existing literature of synchronization in complex networks [5-8]. A synchronization phenomenon has been introduced in a time varying complex dynamical model where synchronizations are estimated with the inner coupling matrix, the eigenvalues and the corresponding eigenvectors of the coupling configuration matrix of the networks [5-6]. A group of synchronized nodes with various levels of synchronization have appeared in a dynamic scale-free network model where most other nodes remain unsynchronized [7]. In another existing literature, A dynamic preferential attachment mechanism has been introduced as synchronization phenomena where dynamic centrality phenomena are reproduced qualitatively to explain real world dynamic networks [8]. In this research paper, we consider two types of synchronizations where we are considering synchronizations of scale-free networks based on the states of nodes with coupling configuration matrix [2-3] and synchronizations of scale-free networks based on the states of nodes with network centralities [12-13].

## II. METHODOLOGY

A network generation algorithm [27] has been implemented where scale-free properties have been followed to generate scale-free networks.

Let, state of node in synchronization of a phase of node $i$ in a general time varying dynamic network model [5],

$$\widehat{x}_i = f(x_i) + \sum_{j=1, j \neq i}^{N} c_{ij}(t)A(t)(x_j - x_i), \quad i = 1, 2, 3, \ldots, N, \quad eq(1)$$

Here, diagonal of coupling configuration matrix [5],

$$c_{ii}(t) = -\sum_{j=1, j \neq i}^{N} c_{ij}(t), \quad i = 1, 2, 3, \ldots, N, \quad eq(2)$$

$eq(1)$ can be written as [5],

$$\widehat{x}_i = f(x_i) + \sum_{j=1, j \neq i}^{N} c_{ij}(t)A(t)x_j, \quad i = 1, 2, 3, \ldots, N, \quad eq(3)$$

Let, G is a Graph with V vertices and E edges where
$G = (V, E), V = \{m_1, m_2, m_3, \ldots, m_n\}$ and $E = \{e_1, e_2, e_3, \ldots, e_n\}$

and $Sd_{ij}(m_k) = $ Shortest distance from node i to node j and $m_k \in$ Sd nodes

1. **Degree Centrality** [10,13]

$$Deg.\ Cen.\ (m_k) = \frac{deg(m_k)}{N-1}, m_k \in V, \quad eq(4)$$

2. **Pagerank Centrality** [10,13]

$$Pag.\ Cen.\ (m_k) = \sum \frac{Pag.\ Cen.\ (m_i)}{deg(m_i)}, m_k \in V, \quad eq(5)$$

3. **Betweenness Centrality** [10,12-13]

$$Bet.\ Cen.\ (m_k) = \frac{2 \sum_{j=1}^{N} \sum_{i=1}^{j-1} \frac{num\ of\ Sd_{ij}(m_k)}{num\ of\ Sd_{ij}}}{N^2 - 3N + 2}, i \neq j \neq k\ and\ m_k \in V, \quad eq(6)$$

4. **Closeness Centrality** [10,13]

$$Clo.\ Cen.\ (m_k) = \frac{N-1}{\sum_{i=1}^{N} Sd(m_i m_k)}, m_k \in V, \quad eq(7)$$

Putting $Deg.\ Cen.\ (m_i)$ of $eq(4)$ in place of $c_{ij}(t)$ of $eq(3)$,

$$\widehat{x}_i = f(x_i) + \sum_{j=1, j \neq i}^{N} Deg.\ Cen.\ (m_i)A(t)x_j, \quad i = 1, 2, 3, \ldots, N, \quad eq(8)$$

Putting $Pag.\ Cen.\ (m_i)$ of $eq(5)$ in place of $c_{ij}(t)$ of $eq(3)$,

$$\widehat{x}_i = f(x_i) + \sum_{j=1, j \neq i}^{N} Pag.\ Cen.\ (m_i)A(t)x_j, \quad i = 1, 2, 3, \ldots, N, \quad eq(9)$$

Putting $Bet.\ Cen.\ (m_i)$ of $eq(6)$ in place of $c_{ij}(t)$ of $eq(3)$,

$$\widehat{x_i} = f(x_i) + \sum_{j=1, j \neq i}^{N} Bet.\ Cen.\ (m_i) A(t) x_j, \quad i = 1, 2, 3, ..., N, \quad eq(10)$$

Putting $Clo.\ Cen.\ (m_i)$ of $eq(7)$ in place of $c_{ij}(t)$ of $eq(3)$,

$$\widehat{x_i} = f(x_i) + \sum_{j=1, j \neq i}^{N} Clo.\ Cen.\ (m_i) A(t) x_j, \quad i = 1, 2, 3, ..., N, \quad eq(11)$$

For experiment, several python libraries which are Networkx [16] version 3.3, Pandas [17] version 2.1.4, NumPy [17] version 1.26.4, google colab [20] version 0.0.1a2, matplotlib [18] version 3.7.1, SciPy [19] version 1.13.1 and others [21-23] have been used in Google colab platform [20]. Documentations have been done with Overleaf, Ubuntu and Google docs [24-26].

## III. RESULTS AND ANALYSIS

Approximately 95% confidence interval (CI) has been considered. Here, only several results along with the worst case scenarios (where phases fluctuated from scale-free properties) have been shown.

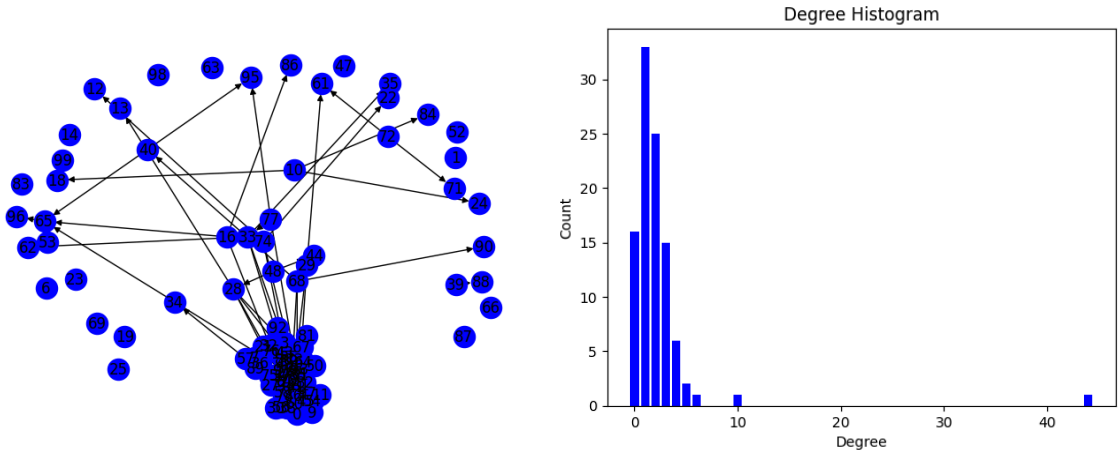

**Figure 1:** one of the phase of dynamic scale-free networks where synchronization is based on "states of nodes"

**Analysis of Figure 1:**

During the time of synchronization of dynamic scale-free networks where synchronization is based on "states of nodes" (see $eq(3)$), nodes with lower degree have a tendency of not connecting with other nodes. For this reason, unconnected nodes appear at the time of synchronization. Because of these unconnected nodes, dynamic scale-free network degree distributions slightly fluctuate from power-law degree distribution.

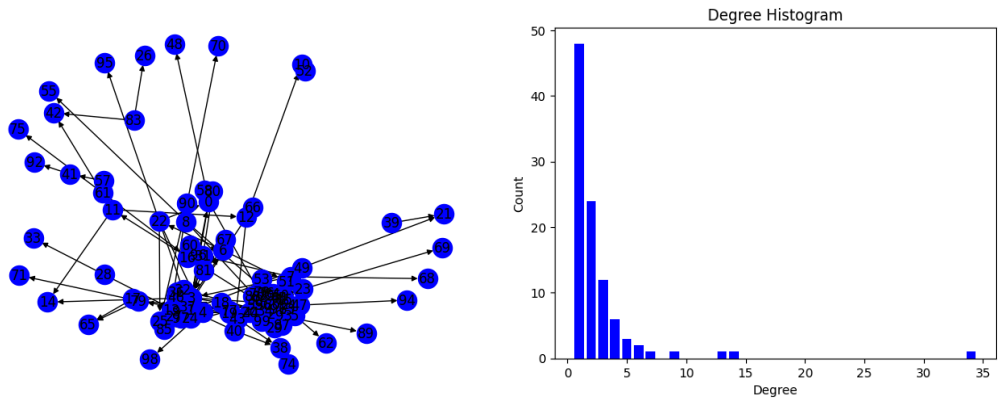

**Figure 2:** one of the phase of dynamic scale-free networks after connecting disconnected nodes

**Analysis of Figure 2:**

To solve these fluctuations from power-law degree distributions, unconnected nodes are connected to a hub in every phase of dynamic scale-free networks. Nodes with one degree are higher than other nodes because scale-free networks have a tendency to produce a higher number of nodes with one degree. Besides that, disconnected nodes are converted into nodes with one degree which makes a higher number of nodes with one degree.

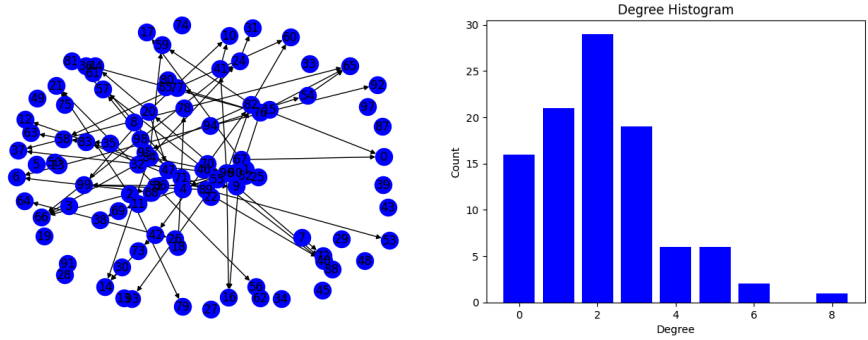

**Figure 3:** one of the phase of dynamic scale-free networks where synchronization is based on closeness centrality

**Analysis of Figure 3 :**

During the time of synchronization based on closeness centrality (see $eq(11)$), some phases's degree distribution of dynamic scale-free networks fluctuate from power-law degree distribution and follow normal degree distribution.

Correlation matrix among results of centrality measures of **Figure 3:**

|  | Degree centrality | Pagerank centrality | Betweenness centrality | Closeness centrality |
| --- | --- | --- | --- | --- |
| Degree centrality | 1.0 | 0.59 | 0.63 | 0.56 |
| Pagerank centrality | 0.59 | 1.0 | 0.4 | 0.77 |
| Betweenness centrality | 0.63 | 0.4 | 1.0 | 0.38 |
| Closeness Centrality | 0.56 | 0.77 | 0.38 | 1.0 |

Table 1: Correlation matrix of centrality measures of Figure 3

From Table 1, we can see, the correlation between pagerank centrality and closeness centrality is 0.77 which is higher than other correlations. It implies important nodes in the correspondent phase of Figure 3 are having tendency to close to most other nodes in the network.

In scale-free networks, result of degree centrality measure and result of betweenness centrality measure are highly correlated (≥0.80) [11]. But, in Table 1, it shows correlation between these two measures is low (0.63) which implies Figure 3 has fluctuated from scale-free properties.

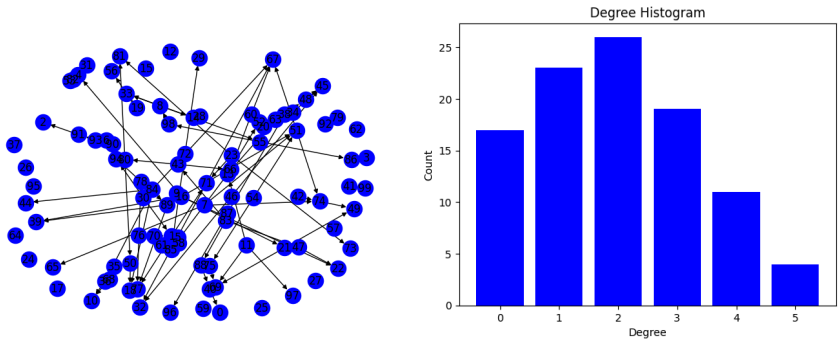

**Figure 4:** one of the phase of dynamic scale-free networks where synchronization is based on betweenness centrality

**Analysis of Figure 4 :**

During the time of synchronization based on betweenness centrality (see $eq(10)$), some phases's degree distribution of dynamic scale-free networks fluctuate from power-law degree distribution and follow normal degree distribution.

Correlation matrix among results of centrality measures of **Figure 4:**

|  | Degree centrality | Pagerank centrality | Betweenness centrality | Closeness centrality |
|---|---|---|---|---|
| Degree centrality | 1.0 | 0.5 | 0.53 | 0.61 |
| Pagerank centrality | 0.5 | 1.0 | 0.37 | 0.89 |
| Betweenness centrality | 0.53 | 0.37 | 1.0 | 0.42 |
| Closeness Centrality | 0.61 | 0.89 | 0.42 | 1.0 |

Table 2: Correlation matrix of centrality measures of Figure 4

From Table 2, we can see, the correlation between pagerank centrality and closeness centrality is 0.89 which is higher than other correlations. It implies important nodes in the correspondent phase of Figure 4 are having tendency to close to most other nodes in the network.

In scale-free networks, result of degree centrality measure and result of betweenness centrality measure are highly correlated (≥0.80) [11]. But, in Table 2, it shows correlation between these two measures is low (0.53) which implies Figure 4 has fluctuated from scale-free properties.

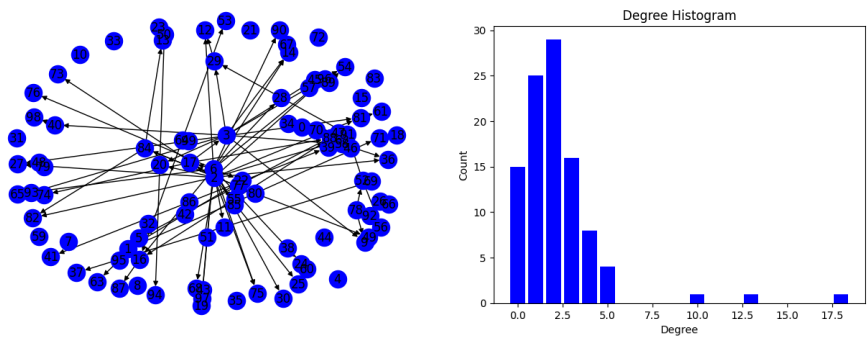

**Figure 5:** one of the phase of dynamic scale-free networks where synchronization is based on pagerank centrality

**Analysis of Figure 5 :**

During the time of synchronization based on pagerank centrality (see $eq(9)$), some phases's degree distribution of dynamic scale-free networks fluctuate from power-law degree distribution and follow slightly normal degree distribution.

Correlation matrix among results of centrality measures of **Figure 5:**

|  | Degree centrality | Pagerank centrality | Betweenness centrality | Closeness centrality |
|---|---|---|---|---|
| Degree centrality | 1.0 | 0.15 | 0.65 | 0.27 |
| Pagerank centrality | 0.15 | 1.0 | 0.64 | 0.85 |
| Betweenness centrality | 0.65 | 0.64 | 1.0 | 0.59 |
| Closeness Centrality | 0.27 | 0.85 | 0.59 | 1.0 |

Table 3: Correlation matrix of centrality measures of Figure 5

From Table 3, we can see, the correlation between pagerank centrality and closeness centrality is 0.85 which is higher than other correlations. It implies important nodes in the correspondent phase of Figure 5 are having tendency to close to most other nodes in the network.

In scale-free networks, result of degree centrality measure and result of betweenness centrality measure are highly correlated ($\geq 0.80$) [11]. But, in Table 3, it shows correlation between these two measures is low (0.65) which implies Figure 5 has fluctuated from scale-free properties.

## IV. Conclusion

Scale-free properties are not always sustainable in synchronization of dynamic scale-free networks. In some of the phases or times in synchronizations of dynamic scale-free networks, scale-free networks fluctuate from its properties and show normally degree distributed dynamic networks. During the time of synchronizations, correlation between degree centrality and betweenness centrality sometimes are not high as it should be high ($\geq 0.80$) for scale-free network properties [11].